\pdfoutput=1   
\documentclass[11pt]{article}
\usepackage[final,numeric,acmfigures]{lucas-arxiv}

\makeatletter
\def\input@path{{./}{../}}
\makeatother
\graphicspath{{./}{../}{figures/}{../figures/}}
\usepackage{multirow}
\usepackage{colortbl}
\venuemacrosbegin   

\usepackage{booktabs}   
\usepackage{tabularx}   
\usepackage{makecell}   
\usepackage{xspace}
\usepackage{xcolor}     
\usepackage{pgfplots}
\pgfplotsset{compat=1.18}
\usetikzlibrary{positioning, arrows.meta, fit, backgrounds, calc, shapes.geometric}

\newcommand{\findingbox}[1]{%
  \noindent\fcolorbox{gray!50}{gray!10}{%
    \parbox{\dimexpr\columnwidth-2\fboxsep-2\fboxrule\relax}{\textbf{#1}}%
  }%
}

\newcommand{\overallbox}[1]{%
	\noindent\fcolorbox{blue!55!black}{blue!6}{%
		\parbox{\dimexpr\columnwidth-2\fboxsep-2\fboxrule\relax}{%
			\textbf{Overall:} #1%
		}%
	}%
}

\venuemacrosend

\logos{\includegraphics[height=\logoheight]{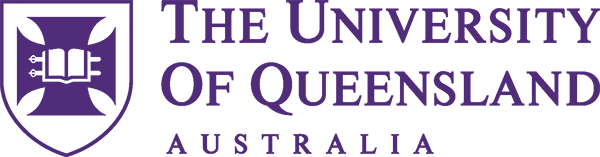},
       \includegraphics[height=\logoheight]{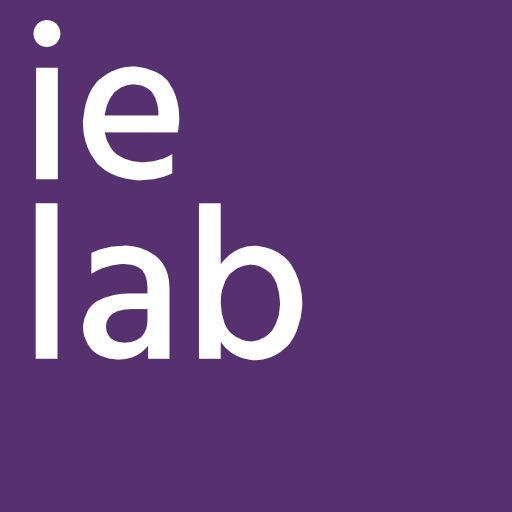}}
\title{Reproducing LightMem: When Does Memory\\Construction Help Over Raw-Turn Retrieval?}
\accepted[10.1145/3848000.3850188]{SIGIR-AP 2026}   
\author{Yongjie Zhou\textsuperscript{1} \and Shuai Wang\textsuperscript{1}
  \and Bevan Koopman\textsuperscript{1,2} \and Guido Zuccon\textsuperscript{1,3}}
\affiliation{\textsuperscript{1}The University of Queensland, Brisbane, Australia
  \quad \textsuperscript{2}CSIRO \quad \textsuperscript{3}Google}
\email{\{yongjie.zhou, shuai.wang2, g.zuccon\}@uq.edu.au, bevan.koopman@csiro.au}
\codelink{https://github.com/ielab/Reproducing-LightMem}
\date{}
\hypersetup{pdftitle={Reproducing LightMem: When Does Memory Construction Help Over Raw-Turn Retrieval?},
  pdfauthor={Yongjie Zhou, Shuai Wang, Bevan Koopman, Guido Zuccon}}

\begin{document}
\maketitle

\begin{abstract}
Long-term conversational agents require access to information from earlier interactions, such as a user's preferences, past requests, or previously mentioned facts. Repeatedly providing the full dialogue history can be expensive as conversations grow, so many memory approaches instead transform past interactions into compact entries that can be retrieved when needed. LightMem is a recent lightweight memory-management approach that reports strong effectiveness while maintaining relatively low construction cost. However, it still relies on a separate constructed memory representation and is evaluated with only one retriever, leaving unclear how sensitive its results are to retriever choice and whether memory construction provides sufficient benefits over retrieving directly from raw turns.

In this study, we reproduce LightMem and compare it with Naive RAG, which retrieves directly from raw user turns. We recover LightMem's main configuration trend, but find that retriever choice substantially affects answer accuracy: changing only the retriever over a fixed LightMem store shifts answer accuracy from 58.1\% to 75.5\%. Retrieval over constructed memory also does not consistently outperform raw-turn retrieval. At matched retrieval depths, Naive RAG generally achieves higher answer accuracy, whereas LightMem has an advantage under tight answering-token budgets. Oracle-reference evaluation further suggests that memory construction may omit or alter answer-relevant information.

Overall, LightMem provides a conditional context-efficiency advantage rather than a consistent answer accuracy advantage over Naive RAG. Strong retrieval over raw turns can match or outperform constructed memory as the answering-token budget increases, motivating future work on retrieval, reranking, query formulation, and their interaction with raw-turn and constructed memory representations. Our code is available at~\url{https://github.com/ielab/Reproducing-LightMem}.

\end{abstract}

\section{Introduction}
\label{sec:intro}

Large language model (LLM) agents require memory to maintain continuity across long-term interactions, such as recalling a user's preferences or earlier requests across conversations. A straightforward approach is to repeatedly provide the full interaction history to the model. However, this becomes increasingly expensive as the history grows and may eventually exceed the model's context window. Memory systems therefore store past interactions and retrieve only the information relevant to the current query.

Many existing systems, including A-MEM~\cite{xu2026mem} and MemoryOS~\cite{kang2025memory}, do not retrieve directly from the original dialogue. Instead, they use LLMs to transform past interactions into structured memory entries by extracting, summarising, linking, or updating information as new interactions arrive. Although these representations may be more compact and organised, constructing them can require hundreds or thousands of LLM calls over a long history, introducing substantial computational cost and latency.

\textbf{LightMem~\cite{fang2025lightmem} is designed to reduce this construction overhead.} It filters tokens from raw dialogue turns and groups related turns by topic, allowing multiple interactions to be processed together during memory construction. This reduces the number of LLM calls compared with systems that construct or update memories for individual interactions.

However, LightMem still transforms the original dialogue into a separate memory representation before retrieval. This transformation introduces an important trade-off. Constructed memories may be shorter and easier to retrieve, but summarising, merging, and updating dialogue turns may also cause \textbf{information loss}. End-to-end answer accuracy alone cannot determine whether a failure arises during \textbf{memory construction} or \textbf{retrieval}. Moreover, LightMem is evaluated using a single retriever, leaving its \textbf{sensitivity to retriever choice} unclear. More fundamentally, it remains uncertain whether \textbf{retrieval over constructed memory outperforms direct retrieval over raw turns}.

We investigate these questions through a reproduction and controlled evaluation of LightMem against a raw-turn retrieval-augmented generation baseline, which we refer to as \textbf{Naive RAG}~\cite{lewis2020retrieval,gao2023retrieval}. We first reproduce LightMem under its reported configurations. We then hold the constructed memory store fixed and evaluate 11 sparse, dense, and hybrid retrievers. Finally, we compare LightMem and Naive RAG using the same retrievers under matched retrieval depths and matched answering-token budgets. We also use oracle-reference conditions to examine potential information loss or alteration during memory construction while reducing the influence of retrieval errors.
We organise the study around the following research questions:

\textbf{RQ1: To what extent can LightMem be reproduced?}
We compare reproduced results with those reported in the original study in terms of answer accuracy and memory-construction efficiency.

\textbf{RQ2: How does retriever choice affect LightMem?}
We hold the memory store and answer-generation settings fixed and evaluate a range of sparse, dense, and hybrid retrieval methods.

\textbf{RQ3: Is memory construction worth it?}
We compare LightMem with Naive RAG under matched retrieval depths and matched answering-token budgets, and use oracle-reference conditions to further examine the effects of memory construction.

Our results recover LightMem's main configuration trend, although its absolute accuracy and construction costs are not fully reproduced. We find that retriever choice substantially affects answer accuracy, while the advantage of constructed memory depends on the answering-token budget. Under tight budgets, LightMem achieves higher answer accuracy for most retrievers, whereas strong retrieval over raw turns often matches or exceeds it as the answering-token budget increases. Oracle-reference comparisons further suggest that memory construction may omit or alter answer-relevant information. These findings suggest that constructed memory provides a conditional context-efficiency advantage rather than a consistent answer accuracy advantage over raw-turn retrieval. They highlight the importance of evaluating memory construction jointly with retrieval choice and answering-token budget, rather than treating memory construction and retrieval as independent components.
\section{Related Work}
\label{sec:related}

\subsection{Memory Systems for LLM Agents}

Memory systems allow LLM agents to retain and reuse information beyond the current context. Early work such as MemGPT~\cite{packer2023memgpt} manages conversation history through a hierarchy of active and external memory. More recent systems, including A-MEM~\cite{xu2026mem} and MemoryOS~\cite{kang2025memory}, transform interactions into structured memory entries that can be linked, updated, and consolidated over time~\cite{chhikara2025mem0,zhong2024memorybank,rezazadeh2025isolated,rasmussen2025zep,tan2025prospect,li2025memos}. Although these systems differ in how memories are represented and maintained, they often rely on repeated LLM calls during memory construction.

A simpler alternative is to store raw dialogue turns and retrieve relevant turns directly at answering time, as in retrieval-augmented generation (RAG)~\cite{lewis2020retrieval,gao2023retrieval}. This avoids a separate construction stage and preserves the original interaction content, but places greater responsibility on the retriever. Our work compares these two designs: retrieval over constructed memory and retrieval over raw dialogue turns.

\subsection{Retrieval Methods}

Sparse retrievers such as BM25~\cite{robertson2009probabilistic} rank documents using lexical overlap, while learned sparse methods such as SPLADE~\cite{formal2021spladev1,formal2021spladev2,lassance2024splade} learn term weights and expansions. Dense retrievers encode queries and documents as vectors and rank them by semantic similarity~\cite{karpukhin2020dense,reimers2019sentence}. Hybrid retrieval combines lexical and dense signals through score fusion~\cite{wang2021bert,cormack2009reciprocal}. Because memory construction changes the wording and granularity of the indexed text, retrievers may behave differently over constructed memories and raw dialogue turns~\cite{pan2025memory}.

\subsection{Evaluating LLM Memory Systems}

LLM memory systems are commonly evaluated through end-to-end answer accuracy on benchmarks such as LongMemEval~\cite{wu2024longmemeval,hu2025evaluating}. However, this metric combines errors from memory construction, retrieval, and answer generation~\cite{xiao2026alpsbench}. An incorrect answer may arise because relevant information was omitted or altered during construction~\cite{chen2025halumem}, missed during retrieval, or not used correctly by the generator~\cite{wang2026memory}. Comparisons are further complicated by differences in retrieval-unit length, since one constructed memory may represent several raw dialogue turns. We address these issues through matched retrieval depths, matched answering-token budgets, and oracle-reference evaluation.
\section{Reproducing LightMem}
\label{sec:lightmem}

LightMem~\cite{fang2025lightmem} is a lightweight long-term memory approach for LLM agents. It compresses and groups dialogue turns before summarising them into long-term memory entries, which are later retrieved to answer new questions. Figure~\ref{fig:lightmem} provides an overview of its sensory, short-term, and long-term memory stages.

\subsection{Memory Construction}

\begin{figure}[t]
	\centering
	\resizebox{0.6\linewidth}{!}{%
	\begin{tikzpicture}[
		font=\small,
		>={Stealth[length=2mm]},
		box/.style={draw, rounded corners=3pt, line width=0.7pt,
			align=center, text width=5.6cm, inner sep=4pt, minimum height=0.7cm},
		io/.style={draw, rounded corners=2pt, fill=gray!12, line width=0.6pt,
			align=center, text width=4.8cm, minimum height=0.45cm, inner sep=3pt},
		flow/.style={-{Stealth[length=2.2mm]}, line width=1pt, gray!70},
		edgelbl/.style={font=\scriptsize\itshape, fill=white, inner sep=1.5pt,
			text=black!75},
		bandlbl/.style={font=\scriptsize\bfseries, rotate=90, anchor=south,
			text=black!60}
	]
		\node[io] (raw) {Raw dialogue turns};

		\node[box, fill=blue!7, draw=blue!55, below=0.4cm of raw] (sensory){%
			\textbf{Sensory memory}\\[1pt]
			\scriptsize LLMLingua-2 token compression (rate~$r$)\\
			\scriptsize $+$ segmentation into topic units};

		\node[box, fill=teal!8, draw=teal!60, below=0.4cm of sensory] (stm){%
			\textbf{Short-term memory}\\[1pt]
			\scriptsize per-topic buffer accumulates compressed turns\\
			\scriptsize until it reaches the capacity threshold~$th$};

		\node[box, fill=orange!9, draw=orange!70, below=0.75cm of stm] (ltm){%
			\textbf{Long-term memory}\\[1pt]
			\scriptsize a batched LLM call turns each buffered segment\\
			\scriptsize into structured memory entries};

		\node[io, cylinder, shape border rotate=90, aspect=0.1,
			fill=orange!14, draw=orange!70, text width=4.0cm, minimum height=0.55cm,
			below=0.6cm of ltm] (store) {Long-term memory store};

		\draw[flow] (raw) -- (sensory);
		\draw[flow] (sensory) -- (stm);
		\draw[flow] (stm) -- node[edgelbl] {triggered once $th$ is reached} (ltm);
		\draw[flow] (ltm) -- (store);

		\draw[flow, dashed] (store.south)
			.. controls +(-0.75,-0.8) and +(0.75,-0.8) .. (store.south);
		\node[edgelbl, align=center] (updlbl) at ($(store.south)+(0,-0.82)$)
			{offline update:\\each new entry revises related existing entries};

		\begin{scope}[on background layer]
			\node[fit=(sensory)(stm)(ltm), draw=black!20, dashed,
				fill=black!3, rounded corners=4pt, inner sep=4pt] (onbg){};
			\node[fit=(store)(updlbl), draw=black!25, dashed,
				fill=black!7, rounded corners=4pt, inner sep=4pt] (offbg){};
		\end{scope}
		\node[bandlbl] at ([xshift=-2pt]onbg.west)
			{Online \scriptsize(interaction)};
		\node[bandlbl] at ([xshift=-2pt]offbg.west)
			{Offline \scriptsize(update)};
	\end{tikzpicture}%
	}
\caption{The original LightMem pipeline. Dialogue turns are compressed and grouped into topic-based segments before being summarised into long-term memory entries. Initial memory construction is performed online, while updates to existing long-term memories are performed offline.}
	\label{fig:lightmem}
\end{figure}

LightMem uses \emph{sensory memory} as the first processing stage for newly arriving dialogue turns. Rather than storing each turn directly, it applies LLMLingua-2~\cite{pan2024llmlingua} to retain only a fraction of its tokens. For a turn containing \(N\) tokens, the compressed turn contains approximately \(N_{\mathrm{comp}} \approx rN\) tokens, where \(r\in(0,1]\) is the compression rate. A larger \(r\) retains more of the original turn and applies less aggressive compression, while a smaller \(r\) removes more tokens. Each compressed turn is then assigned to a topic group and accumulated in the corresponding short-term memory buffer.

Let \(B\) denote the total number of compressed tokens currently stored in a topic buffer. Memory construction is triggered when \(B \geq th\), where \(th\) is the short-term memory capacity threshold. A larger \(th\) allows more related content to accumulate before summarisation, while a smaller \(th\) triggers summarisation earlier and more frequently.

Once the threshold is reached, an LLM summarises the grouped turns into a long-term memory entry. Processing several related turns together avoids constructing a separate memory after every interaction. LightMem also performs an offline update stage that identifies related long-term entries and revises or consolidates them outside the online interaction process.

\subsection{Memory Retrieval and Answering}

At inference time, LightMem uses the current question as a query over the long-term memory store. The original study encodes both the question and each memory entry with the dense sentence-embedding model \texttt{all-MiniLM-L6-v2}~\cite{all_minilm_l6_v2}, ranks the entries by embedding similarity, and selects the top-\(k\) results. The retrieved memories are provided to the answer-generation model together with the current question.

\subsection{Reproduction and Evaluation Framework}
\label{sec:framework}

\begin{figure*}[t]
	\centering
	\includegraphics[width=\textwidth]{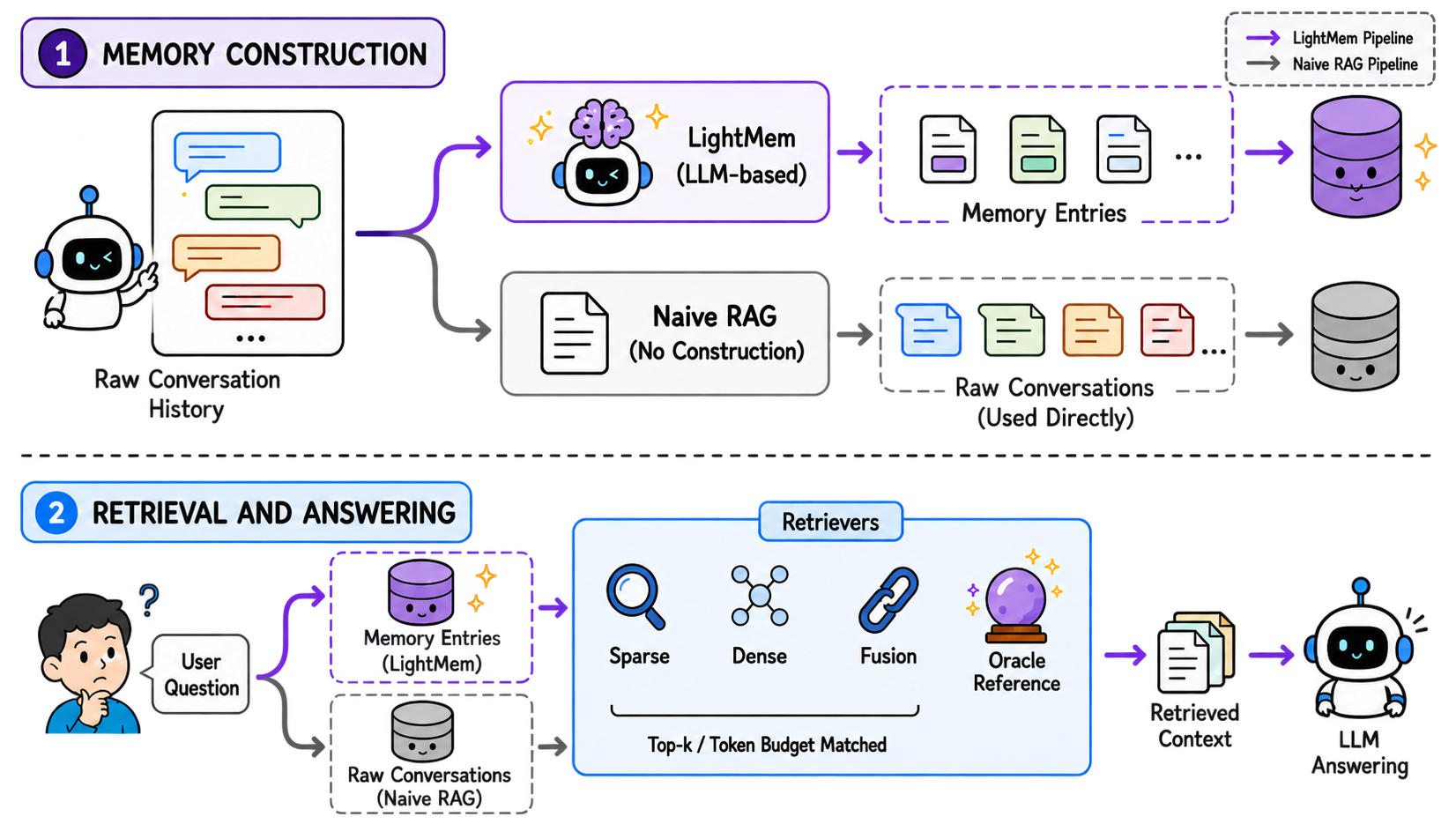}
	\caption{Overview of our evaluation framework. LightMem constructs memory entries from conversation history, whereas Naive RAG retains raw user turns. Both representations are subsequently used for retrieval and answer generation.}
	\label{fig:framework}
\end{figure*}

Our study reproduces the LightMem pipeline and extends its evaluation to examine memory construction and retrieval separately. Figure~\ref{fig:framework} summarises the framework. \textbf{LightMem} transforms raw user turns into constructed memory entries before retrieval, whereas \textbf{Naive RAG} retrieves directly from the raw turns.

We first reproduce the original LightMem evaluation using its reported configurations, default retriever, and comparison baselines, and assess both answer accuracy and construction efficiency. We then fix the constructed memory store and evaluate a broader range of retrievers. Finally, we compare retrieval over constructed memory with retrieval over raw user turns under matched retrievers, retrieval depths, and answering-token budgets. Oracle-reference conditions are used to separate retrieval limitations from potential information loss or alteration during memory construction.

The three research questions share the same dataset, generation model, core LightMem construction settings, and answer-evaluation protocol. Each RQ section therefore specifies only the additional settings required for that experiment.

\par{\textbf{Dataset.}}
We evaluate on LongMemEval-S~\cite{wu2024longmemeval}, a benchmark for long-term conversational memory. It contains 500 questions across six categories: single-session-user, single-session-preference, single-session-assistant, multi-session, knowledge-update, and temporal-reasoning. It also includes abstention questions for which no supporting evidence exists, with gold evidence annotated at the level of raw dialogue turns. Following the original study, we exclude the 56 single-session-assistant questions because LightMem constructs memories only from user turns, leaving 444 questions for all experiments.

\par{\textbf{Generation model.}}
All systems use \texttt{Qwen3-\allowbreak 30B-\allowbreak A3B-\allowbreak Instruct-\allowbreak 2507}~\cite{yang2025qwen3} for answer generation. LightMem also uses the same model for memory summarisation and offline updates. We serve the model locally with vLLM~\cite{kwon2023efficient} (v0.22.1)\footnote{\url{https://github.com/vllm-project/vllm}} and run it on H100 GPUs.

\par{\textbf{LightMem construction settings.}}
The shared LightMem construction settings are reported in Table~\ref{table:hyperparameters}. RQ-specific choices, including the selected \((r,th)\) configuration, retriever, retrieval depth, and comparison condition, are described in the corresponding setup sections.

\begin{table}[t]
  \narrowtable 
	\centering
	\caption{LightMem construction settings used in our reproduction. We evaluate the paired configurations \((r,th)\in\{(0.4, 768), (0.6, 768), (0.8, 1024)\}\); all other settings follow the released implementation.}
	\label{table:hyperparameters}
	\small
	\setlength{\tabcolsep}{4pt}
	
	\begin{tabular*}{\columnwidth}{@{\extracolsep{\fill}}ll@{}}
		\toprule
		Parameter & Value \\
		\midrule
		
		\multicolumn{2}{@{}l}{\textit{Pre-compression and topic segmentation}} \\
		\quad \texttt{pre\_compress}
		& \texttt{true} \\
		\quad compressor
		& \texttt{llmlingua-2} \\
		\quad \texttt{rate} (\(r\))
		& \(\{0.4,0.6,0.8\}\) \\
		\quad \texttt{topic\_segment}
		& \texttt{true} \\
		\quad \texttt{precomp\_topic\_shared}
		& \texttt{true} \\
		
		\midrule
		\multicolumn{2}{@{}l}{\textit{Memory construction}} \\
		\quad \texttt{messages\_use}
		& \texttt{user\_only} \\
		\quad \texttt{metadata\_generate}
		& \texttt{true} \\
		\quad \texttt{text\_summary}
		& \texttt{true} \\
		\quad \texttt{stm\_max\_tokens} (\(th\))
		& \(\{768,1024\}\) \\
		\quad \texttt{max\_tokens}
		& \texttt{16000} \\
		
		\midrule
		\multicolumn{2}{@{}l}{\textit{Offline update (\textsc{OP-update})}} \\
		\quad \texttt{update}
		& \texttt{offline} \\
		\quad \texttt{update\_sim\_threshold}
		& \texttt{0.8} \\
		\quad \texttt{update\_queue\_top\_k}
		& \texttt{20} \\
		\quad \texttt{update\_queue\_keep\_top\_n}
		& \texttt{10} \\
		
		\bottomrule
	\end{tabular*}
\end{table}

\par{\textbf{Answer evaluation.}}
Following LightMem~\cite{fang2025lightmem} and LongMemEval-S~\cite{wu2024longmemeval}, we evaluate generated answers using an LLM judge. We use the fixed \texttt{gpt-5.5-2026-04-23} snapshot\footnote{The original LightMem study uses \texttt{gpt-4o-mini} as the judge. We use a stronger model because we expect it to provide more reliable evaluations. Re-evaluating our outputs with \texttt{gpt-4o-mini} produces similar overall conclusions; these results are reported in Appendix~\ref{app:gpt4o-judge}.} to keep judging consistent across all runs, together with the category-specific evaluation prompts provided in Table~\ref{table:judge_prompt}.\footnote{LongMemEval evaluation prompt source code: \url{https://github.com/xiaowu0162/LongMemEval/blob/main/src/evaluation/evaluate_qa.py}} For abstention questions, an answer is counted as correct only when the model declines to answer.

\begin{table*}[t]
  \centering
  \caption{Category-specific LLM-judge prompts for answer evaluation, taken from LongMemEval~\cite{wu2024longmemeval}. Each prompt returns a yes/no verdict, and placeholders in \texttt{\{...\}} are filled at evaluation time.}
  \label{table:judge_prompt}
  \begingroup
  \footnotesize
  \tcbset{judgecard/.style={enhanced,sharp corners,boxrule=0.5pt,
    colframe=gray!55,colback=gray!3,colbacktitle=gray!18,coltitle=black,
    fonttitle=\normalsize\bfseries,titlerule=0pt,
    left=7pt,right=7pt,top=6pt,bottom=6pt,toptitle=3pt,bottomtitle=3pt,
    fontupper=\footnotesize\raggedright,fontlower=\footnotesize\raggedright,
    segmentation hidden,middle=2pt,space to upper,valign upper=top}}
  \newcommand{\judgefields}[2]{%
    \tcblower
    \texttt{Question: \{question\}}\newline
    #1\newline
    \texttt{Model Response: \{response\}}\newline
    #2%
  }
  \newcommand{\judgeanswerfields}{%
    \judgefields{\texttt{Correct Answer: \{answer\}}}%
      {Is the model response correct? Answer yes or no only.}%
  }

  \begin{tcbraster}[raster columns=2,raster equal height=rows,
      raster column skip=6pt,raster row skip=6pt,raster valign=top]
    \begin{tcolorbox}[judgecard,title={\texttt{single-session-user / multi-session}}]
      I will give you a question, a correct answer, and a response from a model.
      Please answer yes if the response contains the correct answer. Otherwise,
      answer no. If the response is equivalent to the correct answer or contains
      all the intermediate steps to get the correct answer, you should also
      answer yes. If the response only contains a subset of the information
      required by the answer, answer no.
      \judgeanswerfields
    \end{tcolorbox}
    \begin{tcolorbox}[judgecard,title={\texttt{abstention} (any category)}]
      I will give you an unanswerable question, an explanation, and a response
      from a model. Please answer yes if the model correctly identifies the
      question as unanswerable. The model could say that the information is
      incomplete, or some other information is given but the asked information
      is not.
      \judgefields{\texttt{Explanation: \{explanation\}}}%
        {Does the model correctly identify the question as unanswerable? Answer yes
        or no only.}
    \end{tcolorbox}
    \begin{tcolorbox}[judgecard,title={\texttt{knowledge-update}}]
      I will give you a question, a correct answer, and a response from a model.
      Please answer yes if the response contains the correct answer. Otherwise,
      answer no. If the response contains some previous information along with
      an updated answer, the response should be considered as correct as long as
      the updated answer is the required answer.
      \judgeanswerfields
    \end{tcolorbox}
    \begin{tcolorbox}[judgecard,title={\texttt{single-session-preference}}]
      I will give you a question, a rubric for desired personalized response,
      and a response from a model. Please answer yes if the response satisfies
      the desired response. Otherwise, answer no. The model does not need to
      reflect all the points in the rubric. The response is correct as long as
      it recalls and utilizes the user's personal information correctly.
      \judgefields{\texttt{Rubric: \{rubric\}}}%
        {Is the model response correct? Answer yes or no only.}
    \end{tcolorbox}
  \end{tcbraster}

  \vspace{6pt}
  \begin{tcolorbox}[judgecard,title={\texttt{temporal-reasoning}}]
    I will give you a question, a correct answer, and a response from a model.
    Please answer yes if the response contains the correct answer. Otherwise,
    answer no. If the response is equivalent to the correct answer or contains
    all the intermediate steps to get the correct answer, you should also
    answer yes. If the response only contains a subset of the information
    required by the answer, answer no. In addition, do not penalize off-by-one
    errors for the number of days. If the question asks for the number of
    days/weeks/months, etc., and the model makes off-by-one errors (e.g.,
    predicting 19 days when the answer is 18), the model's response is still
    correct.
    \judgeanswerfields
  \end{tcolorbox}
  \endgroup
\end{table*}

\section{RQ1: Can We Reproduce LightMem?}
\label{sec:rq1}

\subsection{Setup}
\label{sec:rq1-setup}

We compare LightMem with the two baselines used in the original study. \emph{Full-context} provides all raw user turns to the answer-generation model without retrieval, while \emph{Naive RAG}~\cite{lewis2020retrieval,gao2023retrieval} retrieves directly from the raw user turns without constructing memory entries.
We evaluate the three reported LightMem configurations:
\[
(r{=}0.4,th{=}768),\quad
(r{=}0.6,th{=}768),\quad
(r{=}0.8,th{=}1024).
\]
Offline updating is enabled in all cases, and the remaining construction settings are listed in Table~\ref{table:hyperparameters}.

For both LightMem and Naive RAG, we use \texttt{all-MiniLM-L6-v2}~\cite{all_minilm_l6_v2}, matching the original study, and retrieve the top-\(10\) retrieval units. The original paper does not report the retrieval depth, so we follow the default setting in the released codebase.

We assess reproducibility using answer accuracy, construction-token usage, and construction LLM calls. Because our judge differs from that used in the original study, we focus on relative method ordering and trends across configurations rather than exact agreement in absolute accuracy.

Construction cost includes online summarisation and offline \textsc{OP-update}. We report construction tokens (\textsc{Constr.}) and construction LLM calls (\textsc{Constr. calls}) as totals over these two stages, which the original study reports separately. Answering-token usage is not reported in the original paper and is therefore included as an additional analysis. Construction costs are measured per ingested sample, while answering tokens are measured per graded question.

\subsection{Results}
\label{sec:rq1-results}

Table~\ref{table:rq1-reproduction} compares our reproduction with the results reported for \texttt{Qwen3-30B-A3B-Instruct-2507}~\cite{yang2025qwen3} in the original LightMem study.

\begin{table*}[t]
  \centering
\caption{Reproduction results on LongMemEval-S. ACC: answer accuracy; Constr.: memory-construction tokens per sample; Constr. calls: construction LLM calls per sample; Ans.: answering tokens per question. Token counts are in thousands; the highest accuracy in each block is in \textbf{bold}.}
  \label{table:rq1-reproduction}
  \small
  \setlength{\tabcolsep}{3.5pt}
  \begin{tabular*}{\textwidth}{@{\extracolsep{\fill}}l ccc c cccc@{}}
    \toprule
    & \multicolumn{3}{c}{Original (from paper)} && \multicolumn{4}{c}{Ours (reproduced)} \\
    \cmidrule(lr){2-4} \cmidrule(lr){6-9}
    Method & ACC (\%) & Constr. & Constr. calls && ACC (\%) & Constr. & Constr. calls & Ans. \\
    \midrule
    Full-context                    & 54.8          & 0     & 0     && 60.8          & 0      & 0      & 18.84 \\
    Naive RAG                       & 60.8          & 0     & 0     && 67.3          & 0      & 0      & 1.03  \\
    LightMem ($r{=}0.4, th{=}768$)  & 62.3          & 144.16 & 192.56 && 57.0          & 72.68  & 64.69  & 0.72  \\
    LightMem ($r{=}0.6, th{=}768$)  & 65.1          & 135.43 & 172.90 && 68.2          & 106.90 & 104.52 & 0.67  \\
    LightMem ($r{=}0.8, th{=}1024$) & \textbf{67.3} & 146.42 & 177.80 && \textbf{70.7} & 119.88 & 117.62 & 0.66  \\
    \bottomrule
  \end{tabular*}
\end{table*}

\vspace{0.5em}
\findingbox{Original finding: LightMem outperforms Full-context and Naive RAG, with accuracy improving across the three reported configurations~\cite{fang2025lightmem}.}
\vspace{0.5em}

\textbf{We reproduce the configuration trend, but not the absolute accuracy values.}
In both studies, accuracy increases across the three configurations, and \((r{=}0.8,th{=}1024)\) performs best. Our results therefore recover the ordering reported in the original study, despite using a different LLM judge.

The comparison with the baselines is less consistent. In our runs, \((r{=}0.6,th{=}768)\) and \((r{=}0.8,th{=}1024)\) achieve 68.2\% and 70.7\% accuracy, respectively, exceeding Full-context at 60.8\% and Naive RAG at 67.3\%. By contrast, \((r{=}0.4,th{=}768)\) reaches 57.0\% and underperforms both baselines. We therefore reproduce LightMem's advantage for the two higher-performing configurations, but not for every reported setting.

\vspace{0.5em}
\findingbox{Reproduction target: LightMem's reported construction-token usage and number of construction LLM calls under the three configurations~\cite{fang2025lightmem}.}
\vspace{0.5em}

\textbf{We reproduce the presence of substantial construction overhead, but not the reported absolute costs.}
The original study reports 135.43--146.42k construction tokens and 172.90--192.56 construction LLM calls per sample. Our corresponding runs require 72.68--119.88k tokens and 64.69--117.62 calls, consistently below the reported values.

The source of this quantitative discrepancy is unclear. Our results nevertheless follow the expected cost pattern: a larger compression rate \(r\) retains more tokens, with \((r{=}0.8,th{=}1024)\) incurring the highest construction cost and \((r{=}0.4,th{=}768)\) the lowest. The original results do not show this ordering as clearly. Construction cost also depends on implementation behaviour, including how many memory entries are created and when offline updates are triggered. Differences in these details may therefore explain the mismatch. Because they are not fully specified in the paper, we follow the released implementation.

Table~\ref{table:rq1-reproduction} additionally reports answering-token usage, which reflects the online cost at answering time. At top-\(10\), LightMem uses 0.66--0.72k answering tokens per question, compared with 1.03k for Naive RAG and 18.84k for Full-context. LightMem therefore reduces the context passed to the answer-generation model, providing an online advantage in answering-token usage. This saving, however, is achieved only after incurring the substantial upfront cost of memory construction.

\vspace{0.5em}
\overallbox{We reproduce LightMem's configuration ordering and show that it can outperform Full-context and Naive RAG with the retriever used in the original study, while also reducing online answering-token usage. However, this advantage does not hold for every configuration, and the reported absolute accuracy and construction-cost values are not reproduced.}
\section{RQ2: How Does Retrieval Affect LightMem?}
\label{sec:rq2}

\subsection{Setup}
\label{sec:rq2-setup}

To isolate the effect of retrieval, we fix the LightMem store constructed with the best-performing configuration from RQ1, \(r{=}0.8\) and \(th{=}1024\), and vary only the retriever used at answering time. Each retriever returns the top-\(10\) memory entries, which are provided to the same answer-generation model.

The original study used only \texttt{all-MiniLM-L6-v2}~\cite{all_minilm_l6_v2}. We extend the evaluation to 11 retrievers from three families:
\begin{itemize}
	\item \textbf{Sparse retrieval:} BM25~\cite{robertson2009probabilistic} provides a lexical baseline, while SPLADE-v3~\cite{lassance2024splade} and PromptReps-sparse~\cite{zhuang2024promptreps} use learned sparse representations.
	
	\item \textbf{Dense retrieval:} We evaluate \texttt{all-MiniLM-L6-v2}, PromptReps-dense, and Qwen3-Embedding~\cite{zhang2025qwen3} at 0.6B, 4B, and 8B scales.
	
	\item \textbf{Hybrid retrieval:} We evaluate PromptReps-hybrid and score-fusion~\cite{wang2021bert} variants combining BM25 with either \texttt{all-MiniLM-L6-v2} or Qwen3-Embedding-0.6B.
\end{itemize}

We report Recall@10 and downstream answer accuracy. LongMemEval-S provides gold evidence at the level of raw dialogue turns rather than constructed memories. We therefore treat a LightMem entry as relevant if it was constructed or updated from a turn marked \texttt{has\_answer}. Recall is computed over the 422 questions with gold evidence; abstention questions are evaluated only by answer accuracy.

We also include an oracle-reference condition in which all memory entries associated with \texttt{has\_answer} turns are provided directly to the answer-generation model. This removes retrieval error under the available relevance mapping, but may introduce irrelevant context because a labelled turn can be split across multiple entries, and individual entries may incorporate information from unlabelled turns. Therefore, this condition is not a strict upper bound on answer accuracy.

\begin{figure*}[t]
	\centering
	\begin{tikzpicture}
		\begin{axis}[
			width=\textwidth,
			height=0.37\textwidth,
			xlabel={Recall@$10$},
			ylabel={Answer accuracy (\%)},
			xmin=0.37, xmax=0.61,
			ymin=56, ymax=79,
			xtick={0.40,0.45,0.50,0.55,0.60},
			ytick={58,62,66,70,74,78},
			grid=both,
			grid style={draw=gray!20},
			tick label style={font=\normalsize},
			label style={font=\large},
			legend columns=3,
			legend style={
				at={(0,1.04)},
				anchor=south west,
				draw=none,
				font=\normalsize,
				column sep=10pt
			},
			clip=false
			]
			
			\addplot[
			only marks,
			mark=square*,
			mark size=3.6pt,
			color=blue!70!black
			] coordinates {
				(0.390,58.1)
				(0.542,71.6)
				(0.456,64.4)
			};
			\addlegendentry{Sparse}

			\addplot[
			only marks,
			mark=*,
			mark size=3.6pt,
			color=green!45!black
			] coordinates {
				(0.530,70.7)
				(0.562,72.7)
				(0.587,75.5)
				(0.583,73.2)
				(0.514,69.1)
			};
			\addlegendentry{Dense}
			
			\addplot[
			only marks,
			mark=diamond*,
			mark size=3.6pt,
			color=purple!75!black
			] coordinates {
				(0.512,69.6)
				(0.502,66.7)
				(0.506,68.2)
			};
			\addlegendentry{Hybrid}
			
			\addplot[
			densely dotted,
			line width=1.4pt,
			color=red!75!black,
			forget plot
			] coordinates {
				(0.37,77.7)
				(0.61,77.7)
			};
			
			\node[
			font=\normalsize,
			fill=white,
			fill opacity=0.95,
			text opacity=1,
			inner sep=2pt,
			anchor=west,
			xshift=6pt
			] at (axis cs:0.390,58.1) {BM25};
			
			\node[
			font=\normalsize,
			fill=white,
			fill opacity=0.95,
			text opacity=1,
			inner sep=2pt,
			anchor=south west,
			xshift=7pt,
			yshift=5pt
			] at (axis cs:0.53,71.6) {SPLADE-v3};
			
			\node[
			name=lbl-qwen4b,
			font=\normalsize,
			fill=white,
			fill opacity=0.95,
			text opacity=1,
			inner sep=2pt,
			anchor=east
			] at (axis cs:0.579,76.1) {Qwen3-Emb-4B};
			\draw[->, gray!55!black, line width=0.7pt, shorten >=2pt, shorten <=1pt]
			(lbl-qwen4b.east) -- (axis cs:0.587,75.5);

			\node[
			name=lbl-qwen8b,
			font=\normalsize,
			fill=white,
			fill opacity=0.95,
			text opacity=1,
			inner sep=2pt,
			anchor=east
			] at (axis cs:0.603,70.9) {Qwen3-Emb-8B};
			\draw[->, gray!55!black, line width=0.7pt, shorten >=2pt, shorten <=1pt]
			(lbl-qwen8b.north) -- (axis cs:0.583,73.2);

			\node[
			name=lbl-promptreps-hybrid,
			font=\normalsize,
			fill=white,
			fill opacity=0.95,
			text opacity=1,
			inner sep=2pt,
			anchor=south east
			] at (axis cs:0.500,70.5) {PromptReps-hybrid};
			\draw[->, gray!55!black, line width=0.7pt, shorten >=2pt, shorten <=1pt]
			(lbl-promptreps-hybrid.south east) -- (axis cs:0.512,69.6);
			
			\node[
			font=\normalsize,
			fill=white,
			fill opacity=0.95,
			text opacity=1,
			inner sep=2pt,
			anchor=south east,
			xshift=-5pt,
			yshift=4pt,
			text=red!75!black
			] at (axis cs:0.608,77.7)
			{Oracle-reference: 77.7\% (Recall@$10{=}1.0$)};
			
		\end{axis}
	\end{tikzpicture}
	
	\caption{Recall@10 versus answer accuracy for LightMem at top-\(10\). Each marker denotes one retriever, and the dotted line marks the oracle-reference accuracy.}
	
	\Description{Scatter plot showing Recall@10 against answer accuracy for 11 retrievers. Points are grouped by retriever family using different colours and markers. Selected retrievers are labelled. A dotted horizontal line at 77.7 percent marks the oracle-reference answer accuracy; its recall of 1.0 lies outside the displayed x-axis range.}
	
	\label{fig:rq2-recall-accuracy}
\end{figure*}

\subsection{Results}
\label{sec:rq2-results}

Figure~\ref{fig:rq2-recall-accuracy} relates Recall@10 to answer accuracy, while Table~\ref{table:rq2-by-category} reports results by question type and significance tests against the default retriever and the oracle-reference condition.

\vspace{0.5em}
\findingbox{Finding 1: Retriever choice substantially affects the effectiveness of LightMem.}
\vspace{0.5em}

With the memory store and answer-generation model fixed, retriever choice produces substantial variation in both Recall@10 and answer accuracy. This shows that LightMem's effectiveness depends strongly on how the same constructed memories are retrieved.

Qwen3-Embedding-4B achieves the highest recall and answer accuracy. Its accuracy of 75.5\% is 2.2 points below the oracle-reference result of 77.7\%, and this difference is not statistically significant. This does not establish equivalence to the oracle-reference condition, but suggests that a strong retriever can recover much of the answerable information available in the fixed memory store. Remaining errors may arise from information omitted or altered during construction or from failures in answer generation.

\begin{table*}[!t]
  \centering
  \definecolor{defrow}{gray}{0.90}
  \providecommand{\rot}[1]{\rotatebox[origin=c]{90}{\textit{\shortstack[c]{#1}}}}
  \caption{Recall@10 / answer accuracy (\%) by question type (\emph{Abst.}: accuracy only). Excluding \emph{Oracle-reference}, the \textbf{best} and \underline{second-best} values in each column are marked, and the \colorbox{defrow}{shaded row} is the default retriever, \texttt{all-MiniLM-L6-v2}. \(^{\mathsf{a}}\)~Significant difference from the default; \(^{\mathsf{b}}\)~significant difference from \emph{Oracle-reference} (accuracy only). Significance testing: paired \(t\)-test (Recall@10) or McNemar's exact test (accuracy) within each column, Bonferroni-corrected, \(p{<}0.05\).}
  \label{table:rq2-by-category}
  \footnotesize\setlength{\tabcolsep}{2.4pt}

  \begin{tabular}{c l ccccc c c}
    \toprule
    & Retriever & Know-upd. & Multi-sess. & Single-pref. & Single-user & Temporal & Abst. & Overall \\
    \midrule
    \multirow{3}{*}{\rot{Sparse}}
      & BM25                & 0.467$^{\mathsf{a}}$ / 76.4 & 0.300$^{\mathsf{a}}$ / 38.0$^{\mathsf{ab}}$ & 0.257$^{\mathsf{a}}$ / 33.3$^{\mathsf{ab}}$ & 0.546$^{\mathsf{a}}$ / 84.4 & 0.382$^{\mathsf{a}}$ / 61.4 & 50.0 & 0.390$^{\mathsf{a}}$ / 58.1$^{\mathsf{ab}}$ \\
      & SPLADE-v3           & 0.595 / \underline{81.9} & 0.475 / 66.1 & 0.641 / 66.7$^{\mathsf{b}}$ & 0.670 / \underline{92.2} & 0.487 / 68.5 & 43.3$^{\mathsf{b}}$ & 0.542 / 71.6 \\
      & PromptReps-sparse & 0.519 / 77.8 & 0.398$^{\mathsf{a}}$ / 46.3$^{\mathsf{b}}$ & 0.333$^{\mathsf{a}}$ / 50.0$^{\mathsf{b}}$ & 0.603 / 89.1 & 0.429 / 67.7 & \underline{53.3} & 0.456$^{\mathsf{a}}$ / 64.4$^{\mathsf{ab}}$ \\
    \cmidrule(l){2-9}
    \multirow{5}{*}{\rot{Dense}}
      & \cellcolor{defrow}all-MiniLM-L6-v2 & \cellcolor{defrow}0.571 / \underline{81.9} & \cellcolor{defrow}0.464 / 59.5 & \cellcolor{defrow}0.607 / 70.0 & \cellcolor{defrow}0.691 / 87.5 & \cellcolor{defrow}0.474 / 71.7 & \cellcolor{defrow}50.0 & \cellcolor{defrow}0.530 / 70.7 \\
      & Qwen3-Emb-0.6B      & 0.600 / \underline{81.9} & 0.520$^{\mathsf{a}}$ / \underline{68.6} & 0.699 / \underline{73.3} & \underline{0.703} / \textbf{93.8} & 0.486 / 67.7 & 43.3 & 0.562$^{\mathsf{a}}$ / 72.7 \\
      & Qwen3-Emb-4B        & \textbf{0.637}$^{\mathsf{a}}$ / \textbf{83.3} & \underline{0.523}$^{\mathsf{a}}$ / \textbf{70.2} & \textbf{0.747} / \textbf{80.0} & 0.699 / \textbf{93.8} & \textbf{0.531}$^{\mathsf{a}}$ / \underline{72.4} & 46.7 & \textbf{0.587}$^{\mathsf{a}}$ / \textbf{75.5} \\
      & Qwen3-Emb-8B        & \underline{0.618} / 80.6 & \textbf{0.532}$^{\mathsf{a}}$ / 66.9 & 0.733 / \underline{73.3} & \textbf{0.712} / 90.6 & \underline{0.510} / \textbf{73.2} & 43.3 & \underline{0.583}$^{\mathsf{a}}$ / \underline{73.2} \\
      & PromptReps-dense & 0.541 / 79.2 & 0.456 / 57.9 & \underline{0.737} / 70.0 & 0.656 / 89.1 & 0.431 / 66.9 & \textbf{56.7} & 0.514 / 69.1$^{\mathsf{b}}$ \\
    \cmidrule(l){2-9}
    \multirow{3}{*}{\rot{Hybrid}}
      & PromptReps-hybrid   & 0.559 / 79.2 & 0.460 / 57.0 & 0.490 / 70.0 & 0.644 / 89.1 & 0.470 / 71.7 & 46.7 & 0.512 / 69.6$^{\mathsf{b}}$ \\
      & MiniLM${+}$BM25     & 0.556 / 80.6 & 0.434 / 54.5 & 0.455$^{\mathsf{a}}$ / 56.7$^{\mathsf{b}}$ & 0.653 / 85.9 & 0.476 / 70.9 & 33.3$^{\mathsf{b}}$ & 0.502$^{\mathsf{a}}$ / 66.7$^{\mathsf{b}}$ \\
      & Qwen3-0.6B${+}$BM25 & 0.566 / 79.2 & 0.451 / 57.9 & 0.421$^{\mathsf{a}}$ / 50.0$^{\mathsf{b}}$ & 0.634 / \underline{92.2} & 0.479 / 70.1 & 43.3$^{\mathsf{b}}$ & 0.506 / 68.2$^{\mathsf{b}}$ \\
    \midrule
    & \emph{Oracle-reference} & \emph{1.000 / 77.8} & \emph{1.000 / 69.4} & \emph{1.000 / 96.7} & \emph{1.000 / 87.5} & \emph{1.000 / 75.6} & \emph{80.0} & \emph{1.000 / 77.7} \\
    \bottomrule
  \end{tabular}
\end{table*}

\vspace{0.5em}
\findingbox{Finding 2: Retriever choice matters most when evidence is distributed across sessions or expressed indirectly.}
\vspace{0.5em}

The largest category-level differences occur for multi-session and single-session-preference questions, and the spread remains substantial even after excluding BM25. This suggests that these categories benefit most from stronger semantic retrieval, likely because the required evidence is distributed across sessions or phrased differently from the query.

Knowledge-update, single-session-user, and temporal-reasoning questions are more stable among non-BM25 retrievers, suggesting that their remaining errors may arise more from memory construction or answer generation than from retrieval. Some retrievers also exceed the oracle-reference result for single-session-user questions, confirming that the \texttt{has\_answer}-based mapping does not capture every useful memory entry.

Abstention questions behave differently because they contain no gold evidence. Retrieved but irrelevant memories may encourage the model to answer when it should abstain.

\vspace{0.5em}
\findingbox{Finding 3: Fusion provides no consistent benefit over the corresponding base retrievers.}
\vspace{0.5em}

The fusion variants do not improve on their corresponding base retrievers. Adding BM25 lowers both recall and answer accuracy for the MiniLM- and Qwen3-based systems, with the recall reduction for MiniLM+BM25 statistically significant relative to the default retriever. PromptReps-hybrid performs only marginally better than PromptReps-dense and does not show a clear end-to-end advantage. Overall, combining lexical and semantic signals provides no consistent benefit over the fixed LightMem store in our evaluation.

\vspace{0.5em}
\overallbox{Retriever choice has a substantial effect on LightMem. Dense and learned sparse retrievers generally perform better than BM25, with the largest benefits observed for multi-session and preference questions, and the best retriever approaching the oracle-reference condition. Combining BM25 with dense retrieval, however, provides no consistent benefit.}
\vspace{0.5em}
\section{RQ3: Is Memory Construction Worth It?}
\label{sec:rq3}
\begin{table}[t]
  \narrowtable 
  \centering
\caption{Retrieval depths paired at matched answering-token budgets. tok/q: mean answering tokens per question, averaged over the 11 retrievers.}
  \label{table:rq3-input-budget-matching}
  \setlength{\tabcolsep}{5pt}
 \begin{tabular*}{\columnwidth}{@{\extracolsep{\fill}}lcccc@{}}
    \toprule
    & \multicolumn{2}{c}{Naive RAG} & \multicolumn{2}{c}{LightMem} \\
    \cmidrule(lr){2-3} \cmidrule(lr){4-5}
    Token budget & top-$k$ & tok/q & top-$k$ & tok/q \\
    \midrule
    $\sim$330 tok/q & 3  & 330 & \textbf{6}  & 329 \\
    $\sim$500 tok/q & 5  & 503 & \textbf{10} & 498 \\
    $\sim$935 tok/q & 10 & 946 & \textbf{20} & 923 \\
    \bottomrule
  \end{tabular*}
\end{table}

\begin{figure*}[t]
	\centering
	  \begin{tikzpicture}[
		legendcell/.style={
			minimum width=0.50cm,
			minimum height=0.28cm,
			draw=white,
			line width=0.4pt
		}
		]
		\node[font=\small\bfseries, anchor=west] at (0,0) {$\Delta$ color:};

		\node[legendcell, fill=blue!35] at (1.8,0) {};
		\node[font=\small, anchor=west] at (2.15,0) {LightMem better};

		\node[legendcell, fill=gray!10] at (5.2,0) {};
		\node[font=\small, anchor=west] at (5.55,0) {near parity};

		\node[legendcell, fill=orange!45] at (7.9,0) {};
		\node[font=\small, anchor=west] at (8.25,0) {Naive RAG better};

		\node[font=\small, anchor=west] at (11.5,0) {darker = larger gap};
	\end{tikzpicture}
	\vspace{1.0em}

	\resizebox{0.87\textwidth}{!}{%
		\begin{tikzpicture}[x=1cm, y=1cm,
			cell/.style={minimum width=1.12cm, minimum height=0.39cm, anchor=center,
				font=\scriptsize, draw=white, line width=0.5pt},
			rowlabel/.style={anchor=east, font=\scriptsize},
			colhead/.style={minimum width=1.12cm, minimum height=0.39cm, anchor=center,
				font=\scriptsize\bfseries, fill=gray!15, draw=white, line width=0.5pt},
			famlabel/.style={rotate=90, anchor=center, font=\scriptsize\itshape}
			]
			\node[font=\scriptsize\bfseries, anchor=east] at (1.25,0) {Retriever};
			\foreach \yy/\name in {
				-0.47/BM25, -0.94/SPLADE-v3, -1.41/PromptReps-sparse,
				-2.57/Qwen3-Emb-0.6B, -3.04/Qwen3-Emb-4B,
				-3.51/Qwen3-Emb-8B, -3.98/PromptReps-dense,
				-4.67/PromptReps-hybrid, -5.14/MiniLM+BM25, -5.61/Qwen3-0.6B+BM25}
			{\node[rowlabel] at (1.25,\yy) {\name};}
			\node[rowlabel, font=\scriptsize\bfseries, fill=gray!25, inner sep=1.6pt, rounded corners=1pt] at (1.25,-2.10) {all-MiniLM-L6-v2};
			\node[famlabel] at (-1.05,-0.94) {Sparse};
			\node[famlabel] at (-1.05,-3.04) {Dense};
			\node[famlabel] at (-1.05,-5.14) {Hybrid};

			\node[font=\footnotesize\bfseries, anchor=center] at (3.265,0.78) {(a) Matched retrieval depth};
			\node[colhead] at (2.12,0) {top-$3$};
			\node[colhead] at (3.30,0) {top-$5$};
			\node[colhead] at (4.48,0) {top-$10$};
			\foreach \yy/\a/\ca/\b/\cb/\c/\cc in {
				-0.47/-7.7^{\dagger}/orange!45/-10.8^{\dagger}/orange!65/-10.6^{\dagger}/orange!65,
				-0.94/-7.0/orange!40/-7.0^{\dagger}/orange!40/-7.2^{\dagger}/orange!45,
				-1.41/-8.6^{\dagger}/orange!50/-6.8^{\dagger}/orange!40/-5.0/orange!30,
				-2.10/+8.8^{\dagger}/blue!45/+5.0/blue!30/+3.4/blue!22,
				-2.57/-3.4/orange!22/-6.5^{\dagger}/orange!38/-5.0/orange!30,
				-3.04/+0.2/gray!10/+1.4/blue!12/-0.5/gray!10,
				-3.51/-1.8/orange!12/-2.9/orange!18/-4.5/orange!28,
				-3.98/+4.3/blue!28/+1.4/blue!12/+1.4/blue!12,
				-4.67/-1.4/gray!10/-7.0^{\dagger}/orange!40/-6.5^{\dagger}/orange!38,
				-5.14/-2.9/orange!18/-8.1^{\dagger}/orange!48/-5.2/orange!32,
				-5.61/-6.8/orange!40/-11.5^{\dagger}/orange!70/-8.8^{\dagger}/orange!52}
			{\node[cell, fill=\ca] at (2.12,\yy) {$\a$};
				\node[cell, fill=\cb] at (3.30,\yy) {$\b$};
				\node[cell, fill=\cc] at (4.48,\yy) {$\c$};}
			\draw[gray!35, line width=0.4pt] (1.47,0.21) rectangle (5.06,-5.85);

			\node[font=\footnotesize\bfseries, anchor=center] at (7.765,0.78) {(b) Matched answering-token budgets};
			\node[colhead] at (6.62,0) {$\sim$330 tok/q};
			\node[colhead] at (7.80,0) {$\sim$500 tok/q};
			\node[colhead] at (8.98,0) {$\sim$935 tok/q};
			\foreach \yy/\a/\ca/\b/\cb/\c/\cc in {
				-0.47/-1.6/orange!10/-4.7/orange!28/-3.2/orange!20,
				-0.94/+4.7/blue!28/-0.2/gray!10/-2.3/orange!14,
				-1.41/-1.1/gray!10/+0.9/gray!10/-2.3/orange!14,
				-2.10/+17.8^{\dagger}/blue!80/+11.3^{\dagger}/blue!55/+7.4^{\dagger}/blue!38,
				-2.57/+2.7/blue!18/-0.7/gray!10/-1.8/orange!12,
				-3.04/+7.4^{\dagger}/blue!38/+5.9/blue!35/+0.7/gray!10,
				-3.51/+5.6/blue!32/+1.8/blue!12/-2.9/orange!18,
				-3.98/+14.6^{\dagger}/blue!70/+12.8^{\dagger}/blue!60/+5.2/blue!32,
				-4.67/+7.2^{\dagger}/blue!38/+0.2/gray!10/-5.0/orange!30,
				-5.14/+5.4/blue!32/-1.1/gray!10/-0.2/gray!10,
				-5.61/-2.3/orange!14/-2.3/orange!14/-6.3^{\dagger}/orange!38}
			{\node[cell, fill=\ca] at (6.62,\yy) {$\a$};
				\node[cell, fill=\cb] at (7.80,\yy) {$\b$};
				\node[cell, fill=\cc] at (8.98,\yy) {$\c$};}
			\draw[gray!35, line width=0.4pt] (5.97,0.21) rectangle (9.56,-5.85);

			\draw[gray!45, line width=0.5pt] (-1.45,-1.755) -- (9.56,-1.755);
			\draw[gray!45, line width=0.5pt] (-1.45,-4.325) -- (9.56,-4.325);
		\end{tikzpicture}%
	}
\caption{Accuracy difference \(\Delta=\mathrm{LightMem}-\mathrm{Naive\ RAG}\) (points) per retriever under (a) matched retrieval depth and (b) matched answering-token budgets. The shaded label marks LightMem's default retriever, \texttt{all-MiniLM-L6-v2}. \(^{\dagger}\)~Significant difference between LightMem and Naive RAG. Significance testing: McNemar's exact test within each column, Bonferroni-corrected, \(p{<}0.05\).}

	\label{fig:rq3-comparison-heatmaps}
\end{figure*}

\subsection{Setup}
\label{sec:rq3-setup}

RQ3 compares two representations of the same conversation history. LightMem retrieves from memories constructed using its best-performing configuration from RQ1, \((r{=}0.8,th{=}1024)\), whereas Naive RAG retrieves directly from raw user turns. Constructed memories provide shorter retrieval units, but may omit details retained in the original dialogue.

We compare the two systems using the same 11 retrievers as in RQ2 under two settings. First, in the \textbf{matched retrieval depth} comparison, both systems use the same retriever and retrieve \(k\in\{3,5,10\}\) retrieval units. This tests whether a constructed memory entry is more useful than a raw turn when the number of retrieved units is fixed.

Second, in the \textbf{matched answering-token budget} comparison (mean input tokens to the answer model per question), we approximately match the context provided to the answer-generation model. Because constructed memories are generally shorter than raw turns, equal retrieval depths give Naive RAG a larger answering context. We therefore measure the mean context length at each depth and select the closest pairs across the two representations. As shown in Table~\ref{table:rq3-input-budget-matching}, LightMem at top-\(6\), top-\(10\), and top-\(20\) approximately matches Naive RAG at top-\(3\), top-\(5\), and top-\(10\), respectively.

Finally, we evaluate an \textbf{oracle-reference condition} for each representation. Naive RAG receives the gold raw turns, while LightMem receives the memories constructed from those turns. Removing retrieval error allows us to test how much answer-relevant information is preserved during memory construction. We also compare LightMem's construction cost with its answering-token savings.

\subsection{Results}
\label{sec:rq3-results}

Figure~\ref{fig:rq3-comparison-heatmaps} compares LightMem and Naive RAG under matched retrieval depths and answering-token budgets. Each cell reports the accuracy difference
\(\Delta=\mathrm{LightMem}-\mathrm{Naive\ RAG}\), such that positive values favour LightMem.

\vspace{0.5em}
\findingbox{Finding 1: When retrieval depth is matched, Naive RAG generally outperforms LightMem.}
\vspace{0.5em}

At the same retriever and retrieval depth, most comparisons favour Naive RAG. Its average advantage is largest at top-\(5\) and remains similar at top-\(10\), showing that constructed memory does not provide a general benefit when both systems retrieve the same number of units.

The result nevertheless depends on the retriever. LightMem consistently performs better with \texttt{all-MiniLM-L6-v2} and PromptReps-dense, while Qwen3-Embedding-4B remains close to parity. These cases suggest that some retrievers interact more favourably with compact memory entries. For most other retrievers, however, retrieving raw user turns is more effective than retrieving its constructed representation.

\vspace{0.5em}
\findingbox{Finding 2: LightMem's average advantage diminishes as the answering-token budget increases.}
\vspace{0.5em}

The matched answering-token budget comparison reveals a different pattern under tight context limits. At the smallest budget, LightMem outperforms Naive RAG for most retrievers, with an average advantage of 5.5 accuracy points. This advantage progressively diminishes as more answering context becomes available and disappears at the largest budget.

LightMem's compact entries allow more retrieved units to fit within a restricted context, improving evidence coverage when the answering-token budget is tight. As more context becomes available, however, Naive RAG can retrieve sufficient raw evidence while preserving the original interaction content.

This benefit remains retriever-dependent. It is most consistent with \texttt{all-MiniLM-L6-v2} and PromptReps-dense, while gains for most other retrievers diminish or reverse at larger budgets. The usefulness of memory construction therefore depends partly on how the retriever interacts with the resulting representation.

\begin{table}
  \narrowtable 
  \centering
\caption{Oracle-reference accuracy and cost of Naive RAG and LightMem. Construction costs are per ingested sample; answering tokens are per question.}
\label{table:rq3-oracle-cost}
  \begin{tabular*}{\columnwidth}{@{\extracolsep{\fill}}lrr@{}}
    \toprule
    Metric & Naive RAG & LightMem \\
    \midrule
    Oracle-reference accuracy (\%) & \textbf{89.0} & 77.7 \\
    Construction tokens & 0 & 119,884 \\
    Construction LLM calls & 0 & 117 \\
    Answering tokens / question & $\sim$903 & $\sim$529 \\
    \bottomrule
  \end{tabular*}
\end{table}

\vspace{0.5em}
\findingbox{Finding 3: Oracle-reference comparisons suggest potential information loss or alteration during memory construction, alongside substantial upfront construction cost.}
\vspace{0.5em}

Table~\ref{table:rq3-oracle-cost} summarises the oracle-reference comparison, construction cost, and answering-token usage. Naive RAG reaches 89.0\% accuracy when given the gold raw turns, compared with 77.7\% for LightMem when given the corresponding constructed memories. With retrieval errors reduced in both conditions, this 11.3-point gap suggests that memory construction may omit or alter answer-relevant information. However, the imperfect evidence mapping prevents us from separating information omission, alteration, and introduced noise. Consistent with this result, Naive RAG with SPLADE-v3 at top-\(10\) reaches 78.8\%, already exceeding the LightMem oracle-reference accuracy. Although the LightMem oracle-reference condition is not a strict upper bound, these results suggest that improving retrieval over the fixed memory store alone may not recover all of the information available in the raw user turns.

LightMem instead reduces online answering-token usage. It lowers the mean answering context from approximately 903 to 529 tokens per question, saving about 374 tokens. Constructing the memory store, however, requires 119,884 tokens and approximately 117 LLM calls per ingested sample. Based on token usage alone, this upfront cost would be recovered only after roughly 321 questions over the same history. This token-only estimate excludes construction latency and monetary costs, as well as embedding, indexing, storage, and retrieval overheads for both systems.

\vspace{0.5em}
\overallbox{Whether memory construction helps over raw-turn retrieval depends on the retriever and the available answering-token budget. With retrievers such as \texttt{all-MiniLM-L6-v2} and PromptReps-dense, compact memories can be beneficial under tight budgets. With stronger retrieval over raw turns or larger answering-token budgets, however, Naive RAG often matches or outperforms LightMem while avoiding LLM-based memory construction and preserving the original interaction content.}

\section{Conclusion and Discussion}
\label{sec:discussion-conclusion}

Our reproduction partially recovers LightMem's reported findings, while our controlled evaluation highlights the importance of retrieval choice in assessing memory construction. Rather than providing a consistent answer accuracy advantage, constructed memory offers conditional context-efficiency benefits that depend on the retriever and available answering-token budget. Strong raw-turn retrieval remains a competitive alternative, particularly as the answering-token budget increases.

Our findings are limited to LightMem evaluated on LongMemEval-S with a single generation model. RQ2 and RQ3 also use a fixed memory construction configuration, while answering-token budgets are matched on average rather than per query. Further evaluation across memory systems, benchmarks, generation models, and construction settings is needed to establish the generality of these findings.

The main lesson is not that memory construction is unnecessary, but that its value cannot be assessed independently of retrieval. Our findings suggest that constructed memories can serve as compact auxiliary representations alongside raw interaction histories, particularly when the available answering context is limited. Preserving the raw interaction history provides a faithful source of evidence, while summaries or structured memories may support efficient access. Future work should study retrieval, reranking, and query formulation together with memory construction, and evaluate construction and retrieval separately so that improvements can be attributed to the correct stage of the system.

\bibliography{../references} 

\appendix
\section{Robustness to the LLM Judge}
\label{app:gpt4o-judge}

The main text judges answers with \texttt{gpt-5.5}, whereas the original
LightMem study~\cite{fang2025lightmem} uses \texttt{gpt-4o-mini}. To test whether
our findings depend on the judge, we re-graded every answer with
\texttt{gpt-4o-mini}, keeping the prompts and all other settings unchanged. Only
answer accuracy changes, while recall, token counts and call counts are unaffected.
The subsections below compare the results under both judges for each research
question.

\subsection{RQ1: Reproduction}
\label{app:gpt4o-rq1}
Table~\ref{table:rq1-judges} compares answer accuracy under both judges. The configuration ordering is unchanged, with \((r{=}0.8,th{=}1024)\) again achieving the highest accuracy among the evaluated configurations and outperforming both Naive RAG and Full-context, while \((r{=}0.4,th{=}768)\) remains below Naive RAG.

\begin{table}[H]
  \centering
  \caption{RQ1 answer accuracy (\%) under the \texttt{gpt-5.5} and \texttt{gpt-4o-mini} judges. The highest accuracy under each judge is in \textbf{bold}.}
  \label{table:rq1-judges}
  \begin{tabular*}{0.65\textwidth}{@{\extracolsep{\fill}}lcc@{}}
    \toprule
    Method & \texttt{gpt-5.5} & \texttt{gpt-4o-mini} \\
    \midrule
    Full-context                    & 60.8 & 56.5 \\
    Naive RAG                       & 67.3 & 66.9 \\
    LightMem ($r{=}0.4, th{=}768$)  & 57.0 & 58.3 \\
    LightMem ($r{=}0.6, th{=}768$)  & 68.2 & 67.8 \\
    LightMem ($r{=}0.8, th{=}1024$) & \textbf{70.7} & \textbf{70.7} \\
    \bottomrule
  \end{tabular*}
\end{table}

\subsection{RQ2: Retriever Choice}
\label{app:gpt4o-rq2}
Retriever choice remains an important factor (Table~\ref{table:rq2-by-category-4omini}). Over the fixed memory store, answer accuracy again varies widely across retrievers. BM25 remains the weakest, while Qwen3-Emb-4B again achieves the highest recall and accuracy.

\begin{table}[H]
  \centering
  \definecolor{defrow}{gray}{0.90}
  \providecommand{\rot}[1]{\rotatebox[origin=c]{90}{\textit{\shortstack[c]{#1}}}}
  \caption{Recall@10 / answer accuracy (\%) by question type under the \texttt{gpt-4o-mini} judge. Highlighting, shading, superscripts and significance testing are as in Table~\ref{table:rq2-by-category}.}
  \label{table:rq2-by-category-4omini}
  \footnotesize
  \setlength{\tabcolsep}{2.4pt}
  \begin{tabular}{c l ccccc c c}
    \toprule
    & Retriever & Know-upd. & Multi-sess. & Single-pref. & Single-user & Temporal & Abst. & Overall \\
    \midrule
    \multirow{3}{*}{\rot{Sparse}}
      & BM25              & 0.467$^{\mathsf{a}}$ / 77.8 & 0.300$^{\mathsf{a}}$ / 38.0$^{\mathsf{ab}}$ & 0.257$^{\mathsf{a}}$ / 33.3$^{\mathsf{ab}}$ & 0.546$^{\mathsf{a}}$ / 82.8 & 0.382$^{\mathsf{a}}$ / 55.1 & \underline{63.3} & 0.390$^{\mathsf{a}}$ / 57.2$^{\mathsf{ab}}$ \\
      & SPLADE-v3         & 0.595 / 81.9 & 0.475 / 66.1 & 0.641 / 73.3 & 0.670 / 89.1 & 0.487 / 62.2 & \underline{63.3} & 0.542 / 71.2 \\
      & PromptReps-sparse & 0.519 / 79.2 & 0.398$^{\mathsf{a}}$ / 47.1$^{\mathsf{b}}$ & 0.333$^{\mathsf{a}}$ / 46.7$^{\mathsf{ab}}$ & 0.603 / 89.1 & 0.429 / 61.4 & \underline{63.3} & 0.456$^{\mathsf{a}}$ / 63.5$^{\mathsf{ab}}$ \\
    \cmidrule(l){2-9}
    \multirow{5}{*}{\rot{Dense}}
      & \cellcolor{defrow}all-MiniLM-L6-v2 & \cellcolor{defrow}0.571 / 79.2 & \cellcolor{defrow}0.464 / 59.5 & \cellcolor{defrow}0.607 / \textbf{86.7} & \cellcolor{defrow}0.691 / 89.1 & \cellcolor{defrow}0.474 / 65.4 & \cellcolor{defrow}\underline{63.3} & \cellcolor{defrow}0.530 / 70.7 \\
      & Qwen3-Emb-0.6B    & 0.600 / \underline{83.3} & 0.520$^{\mathsf{a}}$ / \underline{67.8} & 0.699 / \textbf{86.7} & \underline{0.703} / \textbf{92.2} & 0.486 / 61.4 & 56.7 & 0.562$^{\mathsf{a}}$ / \underline{72.5} \\
      & Qwen3-Emb-4B      & \textbf{0.637}$^{\mathsf{a}}$ / \textbf{84.7} & \underline{0.523}$^{\mathsf{a}}$ / \textbf{69.4} & \textbf{0.747} / \underline{80.0} & 0.699 / \textbf{92.2} & \textbf{0.531}$^{\mathsf{a}}$ / \textbf{68.5} & \textbf{66.7} & \textbf{0.587}$^{\mathsf{a}}$ / \textbf{75.5} \\
      & Qwen3-Emb-8B      & \underline{0.618} / 81.9 & \textbf{0.532}$^{\mathsf{a}}$ / 65.3 & 0.733 / 76.7 & \textbf{0.712} / \underline{90.6} & \underline{0.510} / \underline{66.9} & 60.0 & \underline{0.583}$^{\mathsf{a}}$ / \underline{72.5} \\
      & PromptReps-dense  & 0.541 / 80.6 & 0.456 / 57.0 & \underline{0.737} / \underline{80.0} & 0.656 / \textbf{92.2} & 0.431 / 60.6 & \textbf{66.7} & 0.514 / 69.1 \\
    \cmidrule(l){2-9}
    \multirow{3}{*}{\rot{Hybrid}}
      & PromptReps-hybrid & 0.559 / 80.6 & 0.460 / 57.0 & 0.490 / 66.7$^{\mathsf{b}}$ & 0.644 / 89.1 & 0.470 / 65.4 & 56.7 & 0.512 / 68.5 \\
      & MiniLM${+}$BM25   & 0.556 / 80.6 & 0.434 / 53.7 & 0.455$^{\mathsf{a}}$ / 73.3 & 0.653 / 89.1 & 0.476 / 65.4 & 46.7$^{\mathsf{b}}$ & 0.502$^{\mathsf{a}}$ / 67.3$^{\mathsf{b}}$ \\
      & Qwen3-0.6B${+}$BM25& 0.566 / 79.2 & 0.451 / 57.9 & 0.421$^{\mathsf{a}}$ / 63.3$^{\mathsf{b}}$ & 0.634 / \textbf{92.2} & 0.479 / 63.0 & 56.7 & 0.506 / 68.0$^{\mathsf{b}}$ \\
    \midrule
    & \emph{Oracle-reference} & \emph{1.000 / 76.4} & \emph{1.000 / 68.6} & \emph{1.000 / 96.7} & \emph{1.000 / 90.6} & \emph{1.000 / 66.9} & \emph{83.3} & \emph{1.000 / 75.5} \\
    \bottomrule
  \end{tabular}
\end{table}

\subsection{RQ3: Memory Construction}
\label{app:gpt4o-rq3}
Figure~\ref{fig:rq3-comparison-heatmaps-4omini} shows the same conditional pattern as Figure~\ref{fig:rq3-comparison-heatmaps}. At matched retrieval depths, Naive RAG is more accurate on average, whereas at matched answering-token budgets the average LightMem--Naive RAG difference decreases from \(+4.9\) points at $\sim$330 tok/q to \(-1.3\) points at $\sim$935 tok/q.
The oracle-reference gap also persists, as Naive RAG reaches 87.8\% given the gold raw turns, well above LightMem's oracle-reference accuracy in Table~\ref{table:rq2-by-category-4omini}.

\begin{figure}[!t]
	\centering
	  \begin{tikzpicture}[
		legendcell/.style={
			minimum width=0.50cm,
			minimum height=0.28cm,
			draw=white,
			line width=0.4pt
		}
		]
		\node[font=\small\bfseries, anchor=west] at (0,0) {$\Delta$ color:};

		\node[legendcell, fill=blue!35] at (1.8,0) {};
		\node[font=\small, anchor=west] at (2.15,0) {LightMem better};

		\node[legendcell, fill=gray!10] at (5.2,0) {};
		\node[font=\small, anchor=west] at (5.55,0) {near parity};

		\node[legendcell, fill=orange!45] at (7.9,0) {};
		\node[font=\small, anchor=west] at (8.25,0) {Naive RAG better};

		\node[font=\small, anchor=west] at (11.5,0) {darker = larger gap};
	\end{tikzpicture}
	\vspace{1.0em}

	\resizebox{0.87\textwidth}{!}{%
		\begin{tikzpicture}[x=1cm, y=1cm,
			cell/.style={minimum width=1.12cm, minimum height=0.39cm, anchor=center,
				font=\scriptsize, draw=white, line width=0.5pt},
			rowlabel/.style={anchor=east, font=\scriptsize},
			colhead/.style={minimum width=1.12cm, minimum height=0.39cm, anchor=center,
				font=\scriptsize\bfseries, fill=gray!15, draw=white, line width=0.5pt},
			famlabel/.style={rotate=90, anchor=center, font=\scriptsize\itshape}
			]
			\node[font=\scriptsize\bfseries, anchor=east] at (1.25,0) {Retriever};
			\foreach \yy/\name in {
				-0.47/BM25, -0.94/SPLADE-v3, -1.41/PromptReps-sparse,
				-2.57/Qwen3-Emb-0.6B, -3.04/Qwen3-Emb-4B,
				-3.51/Qwen3-Emb-8B, -3.98/PromptReps-dense,
				-4.67/PromptReps-hybrid, -5.14/MiniLM+BM25, -5.61/Qwen3-0.6B+BM25}
			{\node[rowlabel] at (1.25,\yy) {\name};}
			\node[rowlabel, font=\scriptsize\bfseries, fill=gray!25, inner sep=1.6pt, rounded corners=1pt] at (1.25,-2.10) {all-MiniLM-L6-v2};
			\node[famlabel] at (-1.05,-0.94) {Sparse};
			\node[famlabel] at (-1.05,-3.04) {Dense};
			\node[famlabel] at (-1.05,-5.14) {Hybrid};

			\node[font=\footnotesize\bfseries, anchor=center] at (3.265,0.78) {(a) Matched retrieval depth};
			\node[colhead] at (2.12,0) {top-$3$};
			\node[colhead] at (3.30,0) {top-$5$};
			\node[colhead] at (4.48,0) {top-$10$};
			\foreach \yy/\a/\ca/\b/\cb/\c/\cc in {
					-0.47/-7.0^{\dagger}/orange!42/-10.6^{\dagger}/orange!64/-10.4^{\dagger}/orange!62,
					-0.94/-9.5^{\dagger}/orange!56/-5.6/orange!34/-6.5/orange!39,
					-1.41/-8.3^{\dagger}/orange!50/-8.6^{\dagger}/orange!52/-6.1/orange!37,
					-2.10/+9.2^{\dagger}/blue!55/+5.0/blue!29/+3.8/blue!23,
					-2.57/-4.7/orange!28/-6.5^{\dagger}/orange!39/-6.1/orange!37,
					-3.04/+1.4/blue!12/-1.1/orange!12/+0.7/gray!10,
					-3.51/-1.6/orange!12/-2.3/orange!14/-4.5/orange!27,
					-3.98/+3.4/blue!20/+0.7/gray!10/+2.0/blue!12,
					-4.67/-2.5/orange!15/-6.3/orange!38/-6.8^{\dagger}/orange!40,
					-5.14/-2.0/orange!12/-9.0^{\dagger}/orange!54/-5.4/orange!32,
					-5.61/-7.9^{\dagger}/orange!47/-11.9^{\dagger}/orange!71/-7.9^{\dagger}/orange!47}
			{\node[cell, fill=\ca] at (2.12,\yy) {$\a$};
				\node[cell, fill=\cb] at (3.30,\yy) {$\b$};
				\node[cell, fill=\cc] at (4.48,\yy) {$\c$};}
			\draw[gray!35, line width=0.4pt] (1.47,0.21) rectangle (5.06,-5.85);

			\node[font=\footnotesize\bfseries, anchor=center] at (7.765,0.78) {(b) Matched answering-token budgets};
			\node[colhead] at (6.62,0) {$\sim$330 tok/q};
			\node[colhead] at (7.80,0) {$\sim$500 tok/q};
			\node[colhead] at (8.98,0) {$\sim$935 tok/q};
			\foreach \yy/\a/\ca/\b/\cb/\c/\cc in {
					-0.47/-2.5/orange!12/-5.4/orange!24/-2.0/orange!12,
					-0.94/+1.6/blue!12/+0.7/gray!10/-3.2/orange!14,
					-1.41/-1.4/orange!12/-0.5/gray!10/-1.4/orange!12,
					-2.10/+18.0^{\dagger}/blue!81/+11.0^{\dagger}/blue!50/+6.8/blue!30,
					-2.57/+2.5/blue!12/-0.9/gray!10/-2.7/orange!12,
					-3.04/+9.2^{\dagger}/blue!41/+5.6/blue!26/+2.7/blue!12,
					-3.51/+5.2/blue!23/+1.4/blue!12/-3.6/orange!16,
					-3.98/+13.1^{\dagger}/blue!59/+12.8^{\dagger}/blue!58/+4.3/blue!19,
					-4.67/+5.9/blue!26/+0.0/gray!10/-4.5/orange!20,
					-5.14/+6.1/blue!27/+0.2/gray!10/-3.6/orange!16,
					-5.61/-3.8/orange!18/-2.9/orange!13/-7.2^{\dagger}/orange!32}
			{\node[cell, fill=\ca] at (6.62,\yy) {$\a$};
				\node[cell, fill=\cb] at (7.80,\yy) {$\b$};
				\node[cell, fill=\cc] at (8.98,\yy) {$\c$};}
			\draw[gray!35, line width=0.4pt] (5.97,0.21) rectangle (9.56,-5.85);

			\draw[gray!45, line width=0.5pt] (-1.45,-1.755) -- (9.56,-1.755);
			\draw[gray!45, line width=0.5pt] (-1.45,-4.325) -- (9.56,-4.325);
		\end{tikzpicture}%
	}

	\caption{Accuracy difference \(\Delta=\mathrm{LightMem}-\mathrm{Naive\ RAG}\) (points) per retriever under the \texttt{gpt-4o-mini} judge, for (a) matched retrieval depth and (b) matched answering-token budgets. Colours, shading, \(^{\dagger}\) and significance testing are as in Figure~\ref{fig:rq3-comparison-heatmaps}.}
	\label{fig:rq3-comparison-heatmaps-4omini}
\end{figure}

\section{Exact Accuracies for the RQ3 Comparison}
\label{app:rq3-detailed}

Figure~\ref{fig:rq3-comparison-heatmaps} reports each LightMem--Naive RAG comparison only as the difference \(\Delta\). Table~\ref{table:rq3-detailed} provides the two accuracies underlying each cell for the matched retrieval-depth comparison (panel~(a)) and the matched answering-token-budget comparison (panel~(b)), using the same \texttt{gpt-5.5} judge as in the main text.

\begin{table}[H]
  \centering
  \definecolor{defrow}{gray}{0.90}
  \providecommand{\rot}[1]{\rotatebox[origin=c]{90}{\textit{\shortstack[c]{#1}}}}
  \caption{Answer accuracy behind Figure~\ref{fig:rq3-comparison-heatmaps} for (a) matched retrieval depth and (b) matched answering-token budgets. LM and NR denote LightMem and Naive RAG accuracy (\%), and \(\Delta=\mathrm{LM}-\mathrm{NR}\) (points). Shading, \(^{\dagger}\) and significance testing are as in Figure~\ref{fig:rq3-comparison-heatmaps}.}
  \label{table:rq3-detailed}
  \footnotesize
  \setlength{\tabcolsep}{4pt}
  \newcolumntype{C}{>{\centering\arraybackslash}X}
  \newcolumntype{R}{>{\raggedleft\arraybackslash}X}
  \begin{tabularx}{\textwidth}{@{}c l CCR CCR CCR@{}}
    \toprule
    \multicolumn{11}{l}{\textbf{(a) Matched retrieval depth}} \\
    & & \multicolumn{3}{c}{top-$3$} & \multicolumn{3}{c}{top-$5$} & \multicolumn{3}{c}{top-$10$} \\
    \cmidrule(lr){3-5} \cmidrule(lr){6-8} \cmidrule(lr){9-11}
    & Retriever & LM & NR & \multicolumn{1}{c}{$\Delta$} & LM & NR & \multicolumn{1}{c}{$\Delta$} & LM & NR & \multicolumn{1}{c}{$\Delta$} \\
    \midrule
    \multirow{3}{*}{\rot{Sparse}}
      & BM25 & 47.3 & 55.0 & $-$7.7$^{\dagger}$ & 52.0 & 62.8 & $-$10.8$^{\dagger}$ & 58.1 & 68.7 & $-$10.6$^{\dagger}$ \\
      & SPLADE-v3 & 57.0 & 64.0 & $-$7.0\phantom{$^{\dagger}$} & 64.9 & 71.8 & $-$7.0$^{\dagger}$ & 71.6 & 78.8 & $-$7.2$^{\dagger}$ \\
      & PromptReps-sparse & 50.7 & 59.2 & $-$8.6$^{\dagger}$ & 56.8 & 63.5 & $-$6.8$^{\dagger}$ & 64.4 & 69.4 & $-$5.0\phantom{$^{\dagger}$} \\
    \cmidrule(l){2-11}
    \multirow{5}{*}{\rot{Dense}}
      & \cellcolor{defrow}all-MiniLM-L6-v2 & \cellcolor{defrow}58.8 & \cellcolor{defrow}50.0 & \cellcolor{defrow}+8.8$^{\dagger}$ & \cellcolor{defrow}64.4 & \cellcolor{defrow}59.5 & \cellcolor{defrow}+5.0\phantom{$^{\dagger}$} & \cellcolor{defrow}70.7 & \cellcolor{defrow}67.3 & \cellcolor{defrow}+3.4\phantom{$^{\dagger}$} \\
      & Qwen3-Emb-0.6B & 61.7 & 65.1 & $-$3.4\phantom{$^{\dagger}$} & 66.9 & 73.4 & $-$6.5$^{\dagger}$ & 72.7 & 77.7 & $-$5.0\phantom{$^{\dagger}$} \\
      & Qwen3-Emb-4B & 61.7 & 61.5 & +0.2\phantom{$^{\dagger}$} & 70.9 & 69.6 & +1.4\phantom{$^{\dagger}$} & 75.5 & 75.9 & $-$0.5\phantom{$^{\dagger}$} \\
      & Qwen3-Emb-8B & 60.4 & 62.2 & $-$1.8\phantom{$^{\dagger}$} & 68.5 & 71.4 & $-$2.9\phantom{$^{\dagger}$} & 73.2 & 77.7 & $-$4.5\phantom{$^{\dagger}$} \\
      & PromptReps-dense & 50.9 & 46.6 & +4.3\phantom{$^{\dagger}$} & 57.7 & 56.3 & +1.4\phantom{$^{\dagger}$} & 69.1 & 67.8 & +1.4\phantom{$^{\dagger}$} \\
    \cmidrule(l){2-11}
    \multirow{3}{*}{\rot{Hybrid}}
      & PromptReps-hybrid & 57.9 & 59.2 & $-$1.4\phantom{$^{\dagger}$} & 62.4 & 69.4 & $-$7.0$^{\dagger}$ & 69.6 & 76.1 & $-$6.5$^{\dagger}$ \\
      & MiniLM${+}$BM25 & 55.0 & 57.9 & $-$2.9\phantom{$^{\dagger}$} & 59.7 & 67.8 & $-$8.1$^{\dagger}$ & 66.7 & 71.8 & $-$5.2\phantom{$^{\dagger}$} \\
      & Qwen3-0.6B${+}$BM25 & 56.5 & 63.3 & $-$6.8\phantom{$^{\dagger}$} & 59.0 & 70.5 & $-$11.5$^{\dagger}$ & 68.2 & 77.0 & $-$8.8$^{\dagger}$ \\
    \cmidrule(l){2-11}
    & Mean $\Delta$ & & & $-$2.4\phantom{$^{\dagger}$} & & & $-$4.8\phantom{$^{\dagger}$} & & & $-$4.4\phantom{$^{\dagger}$} \\
    \midrule
    \addlinespace[3pt]
    \multicolumn{11}{l}{\textbf{(b) Matched answering-token budgets}} \\
    & & \multicolumn{3}{c}{$\sim$330 tok/q} & \multicolumn{3}{c}{$\sim$500 tok/q} & \multicolumn{3}{c}{$\sim$935 tok/q} \\
    \cmidrule(lr){3-5} \cmidrule(lr){6-8} \cmidrule(lr){9-11}
    & Retriever & LM & NR & \multicolumn{1}{c}{$\Delta$} & LM & NR & \multicolumn{1}{c}{$\Delta$} & LM & NR & \multicolumn{1}{c}{$\Delta$} \\
    \midrule
    \multirow{3}{*}{\rot{Sparse}}
      & BM25 & 53.4 & 55.0 & $-$1.6\phantom{$^{\dagger}$} & 58.1 & 62.8 & $-$4.7\phantom{$^{\dagger}$} & 65.5 & 68.7 & $-$3.2\phantom{$^{\dagger}$} \\
      & SPLADE-v3 & 68.7 & 64.0 & +4.7\phantom{$^{\dagger}$} & 71.6 & 71.8 & $-$0.2\phantom{$^{\dagger}$} & 76.6 & 78.8 & $-$2.3\phantom{$^{\dagger}$} \\
      & PromptReps-sparse & 58.1 & 59.2 & $-$1.1\phantom{$^{\dagger}$} & 64.4 & 63.5 & +0.9\phantom{$^{\dagger}$} & 67.1 & 69.4 & $-$2.3\phantom{$^{\dagger}$} \\
    \cmidrule(l){2-11}
    \multirow{5}{*}{\rot{Dense}}
      & \cellcolor{defrow}all-MiniLM-L6-v2 & \cellcolor{defrow}67.8 & \cellcolor{defrow}50.0 & \cellcolor{defrow}+17.8$^{\dagger}$ & \cellcolor{defrow}70.7 & \cellcolor{defrow}59.5 & \cellcolor{defrow}+11.3$^{\dagger}$ & \cellcolor{defrow}74.8 & \cellcolor{defrow}67.3 & \cellcolor{defrow}+7.4$^{\dagger}$ \\
      & Qwen3-Emb-0.6B & 67.8 & 65.1 & +2.7\phantom{$^{\dagger}$} & 72.7 & 73.4 & $-$0.7\phantom{$^{\dagger}$} & 75.9 & 77.7 & $-$1.8\phantom{$^{\dagger}$} \\
      & Qwen3-Emb-4B & 68.9 & 61.5 & +7.4$^{\dagger}$ & 75.5 & 69.6 & +5.9\phantom{$^{\dagger}$} & 76.6 & 75.9 & +0.7\phantom{$^{\dagger}$} \\
      & Qwen3-Emb-8B & 67.8 & 62.2 & +5.6\phantom{$^{\dagger}$} & 73.2 & 71.4 & +1.8\phantom{$^{\dagger}$} & 74.8 & 77.7 & $-$2.9\phantom{$^{\dagger}$} \\
      & PromptReps-dense & 61.3 & 46.6 & +14.6$^{\dagger}$ & 69.1 & 56.3 & +12.8$^{\dagger}$ & 73.0 & 67.8 & +5.2\phantom{$^{\dagger}$} \\
    \cmidrule(l){2-11}
    \multirow{3}{*}{\rot{Hybrid}}
      & PromptReps-hybrid & 66.4 & 59.2 & +7.2$^{\dagger}$ & 69.6 & 69.4 & +0.2\phantom{$^{\dagger}$} & 71.2 & 76.1 & $-$5.0\phantom{$^{\dagger}$} \\
      & MiniLM${+}$BM25 & 63.3 & 57.9 & +5.4\phantom{$^{\dagger}$} & 66.7 & 67.8 & $-$1.1\phantom{$^{\dagger}$} & 71.6 & 71.8 & $-$0.2\phantom{$^{\dagger}$} \\
      & Qwen3-0.6B${+}$BM25 & 61.0 & 63.3 & $-$2.3\phantom{$^{\dagger}$} & 68.2 & 70.5 & $-$2.3\phantom{$^{\dagger}$} & 70.7 & 77.0 & $-$6.3$^{\dagger}$ \\
    \cmidrule(l){2-11}
    & Mean $\Delta$ & & & +5.5\phantom{$^{\dagger}$} & & & +2.2\phantom{$^{\dagger}$} & & & $-$1.0\phantom{$^{\dagger}$} \\
    \bottomrule
  \end{tabularx}
\end{table}

\end{document}